\documentclass[12pt,preprint]{aastex}

\usepackage{graphicx}
\shorttitle{An Infrared blast wave in recurrent nova V745 Scorpii }
\shortauthors{Banerjee et al.}
\begin{document}
\title{Near-IR studies of recurrent nova V745 Scorpii during its 2014 outburst}

\author{D. P. K. Banerjee \altaffilmark{1}}
\altaffiltext{1}{Physical Research Laboratory, Navrangpura, Ahmedabad, Gujarat 380009, India}

\author{Vishal Joshi \altaffilmark{1}}
\author{V. Venkataraman \altaffilmark{1}}
\author{N. M. Ashok \altaffilmark{1}}
\author{G.H. Marion\altaffilmark{2,3}}
\author{E.Y. Hsiao\altaffilmark{4}}
\author{A.Raj\altaffilmark{1}}
\altaffiltext{2}{University of Texas at Austin, 1 University Station
C1400, Austin, TX, 78712-0259, USA}
\altaffiltext{3}{Harvard-Smithsonian Center for Astrophysics, 60 Garden
St., Cambridge, MA 02138, USA}
\altaffiltext{4}{Carnegie Observatories, Las Campanas Observatory,
Colina El Pino, Casilla 601, Chile}

\email{orion@prl.res.in}

\begin{abstract}
The recurrent nova (RN) V745 Scorpii underwent its third known outburst on 2014 February 6. Infrared
monitoring of the eruption on an almost daily basis, starting from 1.3d after discovery,
shows the emergence of a powerful blast wave generated by the high velocity nova ejecta exceeding 4000 kms$^{-1}$
plowing into its surrounding environment. The temperature of the shocked gas is raised to a high
value exceeding 10$^{8}$K immediately after outburst commencement. The energetics of the outburst
clearly surpass those of similar symbiotic systems like RS Oph and V407 Cyg which have giant secondaries.
The shock does not show a free-expansion stage but rather shows a decelerative Sedov-Taylor phase
from the beginning. Such strong shockfronts are known to be sites for $\gamma$ ray generation.
V745 Sco is the latest nova, apart from five other known novae, to show $\gamma$ ray emission.
It may be an important testbed to resolve the crucial question whether all novae are generically $\gamma$ ray
emitters by virtue of having a circumbinary reservoir of material that is shocked by the ejecta rather
than $\gamma$ ray generation being restricted to only symbiotic systems with a shocked red giant (RG) wind.
The lack of a free-expansion stage favors V745 Sco to have a density enhancement around the white dwarf (WD),
above that contributed by a RG wind. Our analysis also suggests that the WD in V745 Sco is very massive and
a potential progenitor for a future SN Ia explosion.
\end{abstract}


\section{Introduction}
The symbiotic recurrent nova V745 Scorpii experienced its third known outburst recently
on 2014 February 6.694 UT (Stubbings 2006) with two earlier eruptions being recorded in 1937
and 1989. It is a relatively less well studied nova amongst the 10 currently
known RNe (viz., T Pyx, IM Nor, CI Aql, V2487 Oph, U Sco, V394 CrA, T CrB, RS Oph
and V3890 Sgr) and belongs to the sub-class of RNe
which have giant secondaries (viz., RS Oph, T CrB and V3890 Sgr).
The secondary has been classified to be a giant of spectral type M6 III $\pm$ 2
sub classes (Harrison et al. 1993; Anupama $\&$ Mikolajewska 1999;
Duerbeck et al. 1989; Sekiguchi et al. 1990; Williams et al. 1991) with an
orbital period of 510 $\pm$ 20 days (Schaefer 2009). V745 Sco is a very fast nova
with $t_{2}$ and $t_{3}$ of 6.2 and 9 days respectively which is estimated to lie
at a distance of 7.8 $\pm$ 1.8 kpc in the middle of the galactic bulge (Schaefer 2010).
Optical studies of the 1989 eruption are documented in Sekuguchi et al. (1990) and
Duerbeck et al. (1989). The latter work shows the early spectroscopic evolution through
a montage of 8 spectra covering the period between $\sim$ 10 to 40d after outburst.
Other spectroscopic studies include those by Williams (2003) and Wagner (1989).
In the infrared, Sekiguchi et al. (1990) recorded the lightcurves in the $JHKL$
bands while a near-IR spectrum at $\sim$ 70d after outburst was recorded by Harrison,
Johnson $\&$ Spyromilio (1993). The observational coverage of this RN is sparse and
there are notably no early time IR spectra which record its evolution in the infrared.
This work, and another in preparation, should contribute to filling this gap.

We have been obtaining multi-epoch, photometric and spectroscopic NIR observations in the
0.85 to 2.4 $\mu$m region starting from 1.3d after discovery. During the course of analysis
it was noticed that the emission lines were rapidly narrowing with time. This phenomenon is
rarely seen and is indicative of decelarating matter associated with a shock which in turn
can be associated with $\gamma$ ray generation (see below). The other similar instances
where such a phenomenon was witnessed earlier was in RS Oph (Das et al. 2006) and V407 Cyg
(Munari et al. 2010). Since the development of a strong shock in
V745 Sco is a rare and significant phenomenon, we use part of our data
covering only the {\it I} and {\it J} bands to analyse and study the implications of the event.
In a follow-up paper, we will present a more detailed analysis using our complete $JHK$ spectroscopy
and photometry as in RS Oph (Banerjee et al. 2010).

The present 2014 outburst is being observed at all wavelengths from radio to the
$\gamma$ ray regime
(Rupen et al. 2014, Banerjee et al. 2014, Anupama et al. 2014, Page et al. 2014,
Mukai et al. 2014, Luna et al. 2014, Rana et al. 2014, Cheung et al. 2014).
Among the notable early results is the reported detection of $\gamma$ rays from the
object (Cheung et al. 2014). This has important ramifications for the present study.
$\gamma$ ray detections from nova are recent and few in number and V745 Sco is only
the sixth nova to be detected in $\gamma$ rays after V407 Cyg, Nova Sco 2012, Nova Mon 2012,
Nova Del 2013 and Nova Cen 2013. All detections have been made by the Fermi LAT starting with
the first detection in V407 Cyg in 2010 (Abdo et al. 2010). As per present understanding,
$\gamma$ rays from novae are generated by a diffusive acceleration mechanism as particles
rebound back and forth across a shockfront created by the nova's ejecta plowing into a pre-existing
dense surrounding medium. Generically it is the same principle that leads to creation of
high energy cosmic rays. A shock is thus an essential prerequisite for $\gamma$ ray generation and
it is hence very necessary to establish its presence unambiguously. V407 Cyg was the first $\gamma$-ray
nova where the decelerating shock front was clearly detected (Munari et al. 2010). Unfortunately during
the 2006 RS Oph outburst no $\gamma$-ray observing facility,  with comparable sensitivity as the Fermi telescope,
was available. The generation mechanism of the $\gamma$ ray emission in novae is also under debate.
For the shock to develop it is necessary to have a dense ambient medium
into which the nova ejecta propagates and decelerates. In the case of V407 Cyg and RS oph
the pre-existing dense ambient medium is provided by the high mass-loss from the
secondary late-type giant star (the companion in V407 Cyg is a Mira variable).
In contrast, for instance in the case of Nova Mon 2012 - another $\gamma$ ray nova - it is fairly
certain that the companion is not a late-type giant (Munari et al. 2013) and hence cannot provide
the dense ambient medium through copious mass loss. Thus there is uncertainty as to
how $\gamma$ rays are indeed generated in novae systems. In this context, recent calculations show
that a late-type giant's wind, solely by itself, may not be enough to create the
requisite density enhancements necessary to explain the observed behavior of the $\gamma$ ray
light curve. Additional sources of density enhancement in the form of a reservoir of circum-binary
material around the WD is perhaps needed (Martin $\&$ Dubus 2013). This is a new point of
departure from earlier thinking and interestingly enough, similar arguments for pre-existing circum-binary
material were already proposed by Williams (2013) from totally different considerations.
In such a context, the eruption of V745 Sco is thus an important testbed
for understanding unexplained aspects of $\gamma$ ray generation in novae.

\section{Observations$^{1}$}
\footnotetext[1]{This paper includes data gathered with the 6.5 meter Magellan Telescopes located at Las Campanas Observatory, Chile.}

Near-IR spectroscopy in the
0.85 to 2.4 $\mu$m region at R $\sim$ 1000 was carried out with the 1.2m
telescope of the Mount Abu Infrared Observatory (Banerjee \& Ashok 2012)
using the Near-Infrared Camera/Spectrograph (NICS) equipped with a
1024x1024 HgCdTe Hawaii array. Spectra were recorded with the star dithered to two
positions along the slit with one or more spectra being recorded in both
of these positions. The coadded spectra in the respective dithered
positions were subtracted from each other to remove sky and dark
contributions. The spectra from these sky-subtracted images were
extracted  using IRAF tasks  and wavelength calibrated using a combination
of OH sky lines and telluric lines that register with the stellar spectra.
To remove telluric lines from the target's spectra, it was ratioed with the
spectra of a standard star from whose spectra
the Hydrogen Paschen and Brackett absorption lines had been removed.
The spectra were finally multiplied by a blackbody at the effective
temperature of the standard stars SAO 186061 \& SAO 209303. The log of the
observations is given in Table 1.

\begin{figure}
\includegraphics[bb = 176 91 470 655, width = 3.5in, clip]{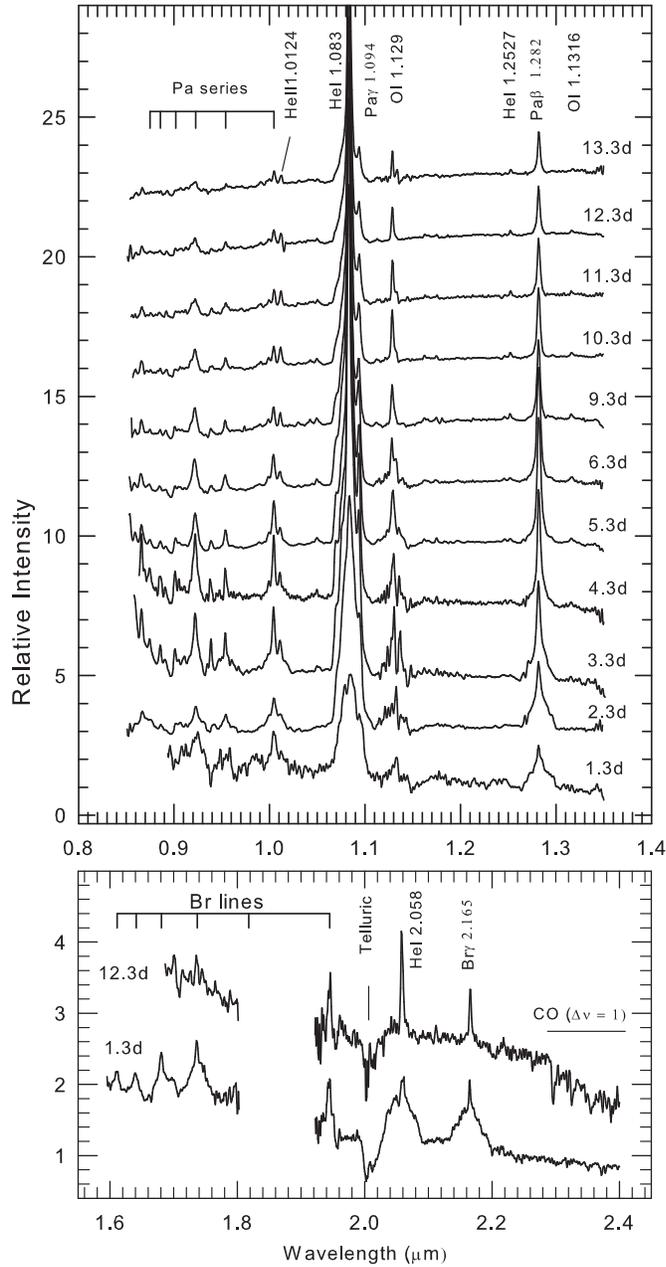}
\caption{Temporal evolution of the IJ spectra (top panel) with days after
outburst indicated. Representative HK
spectra are shown (bottom panel) on 1.3d \& 12.3d after outburst.The prominent lines are
marked.}
 \label{fig1}
\end{figure}

\begin{figure}
\includegraphics[bb = 216 207 531 638, width = 3.5in, clip]{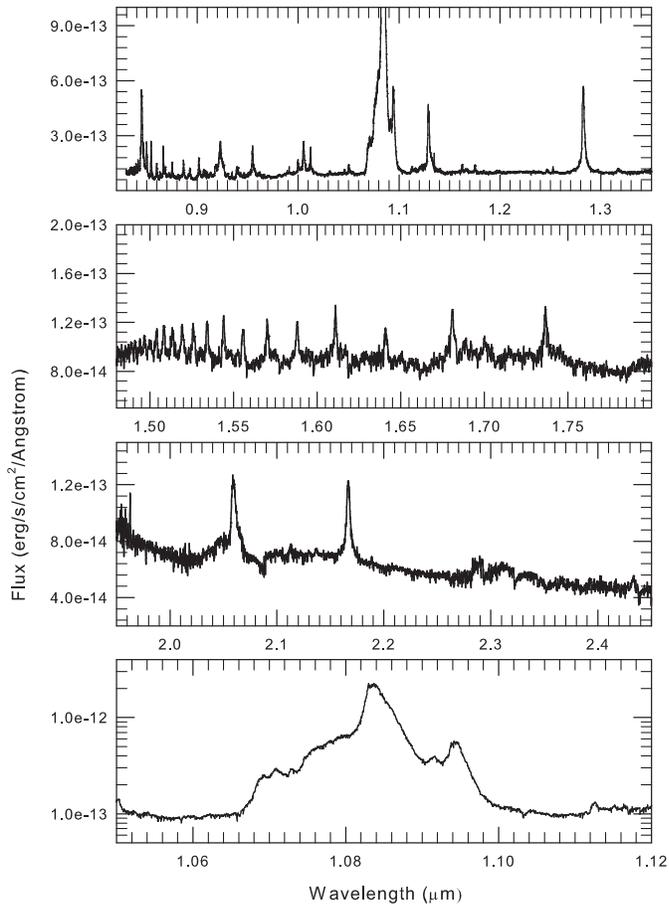}
\caption{ The 0.8-2.5 $\mu$m FIRE spectrum  on February 15.4 (8.7d after outburst)
is shown in top three panels. The bottom panel shows an expanded view of the HeI 1.0830 $\mu$m line.}
\end{figure}

\begin{figure}
\epsscale{0.6}
\rotatebox{0}{\plotone{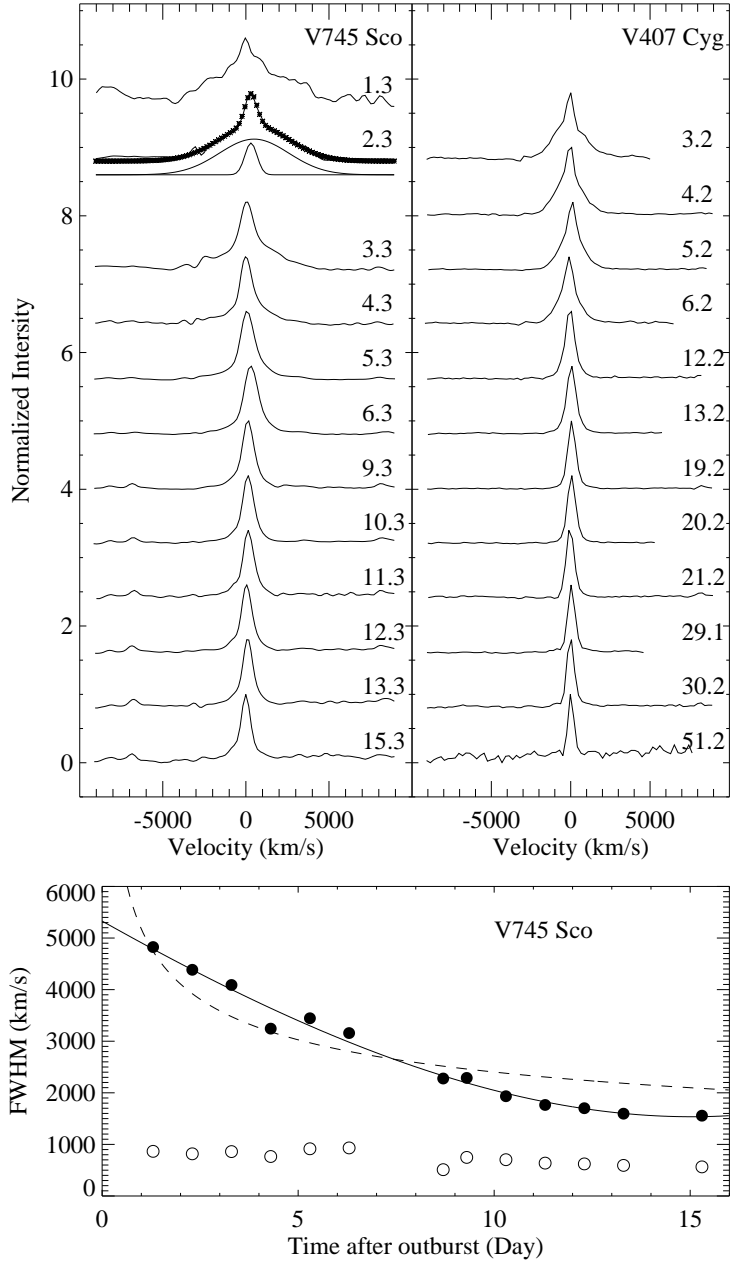}}
\caption{Pa$\beta$ line profiles showing narrowing of the line width in V745 Sco (top left panel; details in section 3).
A sample two gaussian fit to the profile 2.3d after outburst is shown. The narrow and broad components are shown and their added sum is shown by star symbols.
Similar plots for the line narrowing in the 2010 outburst of V407 Cyg are shown in the top right panel.
The lower panel shows the temporal evolution of the FWHM  of the broad and narrow components
(dark and empty circles respectively); the velocity decline is poorly fitted by a t$^{-1/3}$ law (dashed lines)
but better approximated by a third degree polynomial (continuous line)
}
\end{figure}

\begin{table}[ht]
\caption{Log of observations}
\centering
\begin{tabular}{c c c c c}
\hline
UT & Days after & Airmass & Telluric & Airmass \\
& outburst & V745 Sco & Standard & Standard \\
\hline
2014 Feb 08.0 & 1.3 & 2.58 & SAO186061 & 2.05 \\
2014 Feb 09.0 & 2.3 & 3.29 & SAO186061 & 2.51 \\
2014 Feb 10.0 & 3.3 & 2.97 & SAO186061 & 2.03 \\
2014 Feb 11.0 & 4.3 & 3.12 & SAO186061 & 2.16 \\
2014 Feb 12.0 & 5.3 & 2.91 & SAO186061 & 2.08 \\
2014 Feb 13.0 & 6.3 & 2.73 & SAO186061 & 2.01 \\
2014 Feb 15.46$^{a}$ & 8.7 & 1.33 & HD163084 & 1.33 \\
2014 Feb 16.0 & 9.3 & 3.06 & SAO209303 & 3.13 \\
2014 Feb 17.0 & 10.3 & 3.06 & SAO209303 & 3.13 \\
2014 Feb 18.0 & 11.3 & 2.54 & SAO209303 & 2.53 \\
2014 Feb 19.0 & 12.3 & 3.05 & SAO209303 & 3.12 \\
2014 Feb 20.0 & 13.3 & 2.50 & SAO209303 & 2.54 \\
2014 Feb 22.0 & 15.3 & 2.33 & SAO209303 & 2.14 \\ [1ex]
\hline
\tablenotetext{a}{Observation from Magellan South. Remaining observations are from Mt. Abu.
The outburst date is taken to be the discovery date viz. 2014 Feb 6.7}
\end{tabular}
\end{table}

NIR spectra ($R \approx 6000$, $\lambda = 0.8 - 2.5$$\mu$m) were obtained
using the FoldedPort Infrared Echellette (FIRE) spectrograph on the 6.5m
Magellan Baade Telescope (Simco et al. 2008). FIRE data are reduced using a
custom IDL package (Simco et al. 2008) and additional
procedures described by Hsiao et al. 2013.

\section{Results and Discussion}
We present in Figure 1 the complete set of spectra in the $\it IJ$ band between 0.85 to 1.35 $\mu$ms
obtained from Mount Abu. The major lines are marked and it may be seen that most of the lines
are of H and He with HeI 1.0830 $\mu$m overwhelming all other lines in strength. The NIR spectra
are typical of the He/N class of novae (Banerjee \& Ashok 2012) and
are fairly similar to those observed in RS Oph except that the strength of Lyman beta
fluoresced OI 1.1287 $\mu$m line develops relatively much slower in V745 Sco. Unlike
Duerbeck et al. (1989), who detected coronal lines $\sim$ 10d after outburst, we do not
see any coronal lines during the span of our observations. The highest excitation lines
seen here are due to HeII. The spectrum obtained from FIRE is showed in Figure 2.
All the emission lines of Figure 1 are clearly seen here too but at higher resolution.
The sequence of Brackett lines in the H band between Br 10 at 1.7362 $\mu$m to Br 25 at 1.4967 $\mu$m is rather striking.
A magnified view shows that mild first overtone CO emission at 2.29 $\mu$m and beyond,
arising from the secondary, had already begun to appear in the FIRE spectrum.
The CO features became more pronounced with time as emission from the secondary becomes
dominant. This is demonstrated in the bottom panel of Figure 1 which shows two $\it K$ band spectra;
one on 1.3d and the other 12.3d after outburst-- in the latter the CO bands are clearly seen.

Figure 3 well illustrates the evolution of the profile of
the Pa$\beta$ 1.2818 $\mu$m line which was chosen since it
is both a strong line and also unblended with other lines.
 Profile width measurements are therefore reliable.
 The observed profiles are composed of a broad component which is attributed to the nova ejecta
 on which is superposed a sharp and narrow component. The profiles are very similar to those
 seen in V407 Cyg in which the sharp component was attributed
 to the sudden ionization of a large fraction of the secondary's wind by
the flash of energetic radiation produced by the thermonuclear event (Munari et al. 2010).
A similar origin is proposed here too. What is striking is the rapid narrowing of the profiles
with time. We decomposed each profile into
two gaussians representing the broad and narrow components respectively
and measured the evolution of their FWHM (full width at half-maximum) values. A representative two gaussian fit to
one of the profiles is shown in Figure 3 with the variation of the FWHM's with time
is shown in the bottom panel.
The right panel shows a similar evolution for the 2010 outburst of V407 Cyg from unpublished material
not included in Munari et al (2010). The shape and evolution of the profiles of both objects share a
good similarity. For V745 Sco, the narrow component from the Mira wind does not show much variation.
During the early stages it is kinematically perturbed to some extent but its FWHM gradually tends
to evolve towards its value observed at quiescence. We consider the FWHM of the
H lines in quiescence to be adequately represented by the FWHM of the H$\beta$ line whose intrinsic width (corrected for instrumental broadening) is $\sim$ 450 $\pm$ 15 kms$^{-1}$ (Munari, private communication) as presented in the high-resolution atlas of symbiotic stars by
Munari \& Zwitter (2002). We discuss the evolution of the broad
component in V745 Sco below.

The behavior of the shockwave as it propagates into the dense ambient medium surrounding the WD
is usually divided
into the following stages (e.g. Bode \& Kahn 1985). First is a free expansion or ejecta-dominated stage,
where the ejecta expands freely and the shock moves at a
constant speed without being impeded by the surrounding
medium. This phase generally extends to the time it takes for the
 swept-up mass to equal the ejecta mass. The second phase is a Sedov-Taylor stage,
where the majority of the ejecta kinetic energy has been transferred
to the swept-up ambient gas. This is an adiabatic phase since the shocked material is so hot that
there is negligible cooling by radiation losses. During this phase a deceleration is seen in
the shock whose velocity {\it v} versus time {\it t} is expected to behave
as {\it v} $\propto$ $t^{-1/3}$, assuming a $r^{-2}$ dependence for the decrease in
density of the wind. In phase 3, the shocked material has cooled
by radiation, and here the expected dependence of the shock
velocity is {\it v} $\propto$ $t^{-1/2}$.  One may mention that the strong X-ray blast wave,
seen during the 2006 outburst of RS Oph, largely conformed to the above behavior (Sokolski et al. 2006; Bode et al. 2006).

The free expansion stage was not seen in V745 Sco whose implications need to be understood. The intrinsic colors of the late
M type giant in V745 Sco are estimated to be $(J-H)$$_{0}$ $\sim$0.95 and $(H-K)$$_{0}$ $\sim$0.41 using 2MASS magnitudes
of $JHK$ = 10.04, 8.85 $\&$ 8.3 respectively as quiescent values and correcting them using an adopted reddening value
of $E(B-V)$ = 0.70 (Schlafly \& Finkbeiner 2011). A value of $E(B-V)$ $>$ 0.6, for the estimated distance of 7.8 kpc to V745 Sco,
is also supported from modelling of the galactic interstellar extinction by Marshall et al. (2006).
On an IR color-color diagram (see Figs 6 \& 4 of Whitelock \& Munari, 1992) this places it
among galactic bulge giants, a conclusion that was also
 reached by Sekiguchi et al. (1990). A comparison by Whitelock \& Munari (1992) of
 the IR characteristics of neighborhood M
giants, bulge M giants and the M giants of S type symbiotic systems shows that the M
giant secondaries in symbiotic systems
are very similar to those in the bulge and are thus low mass ($<$ 1 M$_{\odot}$) objects.
We adopt this as an upper limit for the mass of the secondary in V745 Sco.  It is also likely
that a significant fraction of all the symbiotic
M stars are actually asymptotic giant branch (AGB) stars rather than giant branch stars (Kniazev et al. 2009).
This suggestion that symbiotics have AGB stars as mass donors would support the
view that additional mass could be transferred through the stellar
winds, above that transferred via Roche lobe overflow, since the winds
from AGB stars are stronger than from normal giants. We thus adopt a mass loss
rate of around 10$^{-7}$ M$_{\odot}$yr$^{-1}$. This is a reasonable value for an AGB
star and also in line with that
chosen for V407 Cyg (Martin $\&$ Dubus; 2013).

Assuming a high-mass for the WD
(in the range 1.2 to 1.4 M$_{\odot}$), M$_{sec}$ $\sim$ 1M$_{\odot}$ and orbital
 period of 520d, the separation
between the binary components in V745 Sco is tightly constrained in the range 1.4 to 1.5 AU.
In the elapsed time of 1.3d between onset of outburst and
our first observation, the blast wave traveling at at over 4000 kms$^{-1}$ will sweep up material
within a radius of 3 AU from the WD.
The mass of this material is estimated to be M$_{swept}$ = 0.7$\times$10$^{-7}$M$_{\odot}$
assuming (d$\dot{\textrm{M}}$/dt)$_{secondary}$ = 10$^{-7}$ M$_{\odot}$yr$^{-1}$, a geometric 1/r$^{2}$
dilution in the density profile of the RG wind and a velocity of the RG wind of 10 km/s.
Increasing the wind velocity will reduce M$_{swept}$ while
increasing (dM/dt)$_{secondary}$ will linearly increase it. But for a reasonable physical
choice of parameters M$_{swept}$ appears constrained between
10$^{-7}$ to 10$^{-6}$ M$_{\odot}$. M$_{swept}$ is estimated assuming the material
between WD and secondary has no additional enhancements beyond that due to spherically
symmetric wind from the secondary (see Figure 1 of Martin $\&$ Dubus, 2013). The free
expansion stage, if it ever occurred, had commenced and completed before our first
observation made 1.3d after discovery. The small value of matter swept out
during these 1.3d indicates one of two possibilities. First, the mass of the
ejected matter M$_{ej}$ in the outburst is small
and of the order of M$_{swept}$. The small value of M$_{ej}$ in turn would imply that the
central WD is massive since the
critical mass of the accreted envelope required to trigger
a thermonuclear runaway is inversely proportional to the mass of the WD

($M_{acc}=\frac{4\pi{R_{WD}}^4P_{crit}}{GM_{WD}}$ where $P_{crit}$=1020 dyne
cm$^{-2}$ for $M_{WD}$ = 1.4 $M_{\odot}$; Truran $\&$ Livio 1986). The second
conclusion that can be drawn from a free expansion stage that is either very
short-lived (< 1.3 d) or absent is that the ejecta was very quickly impeded by
additional material apart from the giant's wind. That is, the Martin \& Dubus (2013)
hypothesis positing additional material enhancement, as applicable to V407 Cyg,
is valid here too. Preliminary results from the $\gamma$ ray detection by
Fermi-LAT data indicates the detections with largest observed significances were
on 2014 February 6 and 7 with no significant emission (within stipulated detection
limits given) was detected in the subsequent days through the end of 2014 February 10.
The fact that the $\gamma$ rays peaked early, coincident with the optical outburst,
strongly points at a dense circumbinary reservoir around the WD. The greatest
difficulty that Martin \& Dubus (2013) faced while reproducing the $\gamma$ ray
lightcurve of of V407 Cyg was in simulating the early peaking of the $\gamma$ ray
emission using just a RG wind. To overcome this they were forced to invoke the
presence of dense additional matter close to the WD. It may be noted that the
above argument does not rule out the possibility of a small ejecta mass
(or equivalently a massive WD). Independent support for a high mass WD is found in
the extremely early turn-on of the super-soft X-ray phase at $\sim$3d after discovery
(Page et al. 2014). This, coupled with the very high ejecta velocities observed,
 imply the presence of very low mass ejecta and thereby a massive WD (Figure 6 of Schwarz et al., 2011).
  V745 Sco thus could be a potential progenitor candidate
 for a SN Ia explosion by virtue of having a high mass WD whose mass additionally, as in other similar symbiotic systems like RS Oph,
 T CrB and V3890 Sgr, is suggested to be increasing after each outburst (Hachisu $\&$ Kato 2001; Hachisu, Kato $\&$ Luna 2007).

The adiabatic (decelerative) phase in Figure 2 deviates significantly from the $t^{-1/3}$ dependency
indicating that the shock is propagating into a wind which is not spherically symmetric. The reasons
for this are two fold. The nova shell is expected to be slowed down more effectively in the parts
moving in the direction of the RG due to the increasing density in that direction. In addition,
as discussed above, there is additional material, most likely distributed
over the equatorial plane. The combined effect of these is to make matter distribution around the
WD's position anisotropic and the shockfront should thus rapidly becomes aspherical.
 Good support for this is offered by the detailed radio monitoring and modeling of the V407 Cyg outburst by
Chomiuk et al. (2012; refer their Figure 6). V745 Sco was observed 10d after the optical discovery
on February 16 with NuSTAR showing a luminous hard X-ray source whose spectrum could be
modeled by a plasma in collisional ionization equilibrium at kT = 2.6 keV or equivalently
3.02$\times$10$^{7}$K (Rana et al. 2014). This is consistent with what we observe. For a strong shock, the
post-shock temperature
$T_{s}$ is given by $T_{s}= \frac{3\bar{m}v^{2}}{16k}$ where k is the Boltzmann constant
and $\bar{m}=10^{-24}$g is the mean particle mass including electrons (Bode et al. 2006).
On day 10.3 we measure $v$ = 1930 kms$^{-1}$ equivalent to a temperature of 5.05$\times$10$^{7}$K which
agrees satisfactorily with the X ray result. Extending the calculations, the gas must have been
heated to extremely high temperatures exceeding 1$\times$10$^{8}$K at 1.3d when the FWHM was
4825 kms$^{-1}$. In comparison, in RS Oph whose evolution was very well documented, a Pa$\beta$ FWHM
of 3066 kms$^{-1}$ on 1.16d was measured (Das et al. 2006) and for the same line in V407 Cyg we
obtained a value 1862 kms$^{-1}$ on 2010 March 13 (3.2d after outburst; Munari et al. 2010
measured a FWHM of 2760 kms$^{-1}$ on day +2.3 from the H $\alpha$ profile). Clearly the outburst of
V745 Sco is extremely powerful, an aspect that needs to be emphasized and which was not
established from its earlier outbursts. Its energetics overshadow even those of RS Oph and V407 Cyg.

Research at PRL is supported by the Department of Space, Government of India. GHM thanks D. Osip, P. Palunas,
Y. Beletsky and the engineering group at the Las Campanas Observatory for their support of observations.
The CfA Supernova Program is supported by NSF grant AST-1211196 to the Harvard College Observatory.
We thank the reviewer Prof Ulisse Munari for helpful comments.


\end{document}